\documentclass[letterpaper]{article}
\usepackage{aaai2026}
\usepackage{times}
\usepackage{helvet}
\usepackage{courier}
\usepackage[hyphens]{url}
\usepackage{graphicx}
\usepackage{natbib}
\usepackage{caption}
\usepackage{amsmath}
\usepackage{amssymb}
\usepackage{amsthm}

\newcommand{\eps}{\varepsilon}

\newcommand{\gstar}{\gamma_t^{\star}}
\newcommand{\bstar}{B/Y_t}
\newcommand{\alphabar}{\bar\alpha}
\newcommand{\kappabar}{\bar\kappa}

\def\includeappendix{1}
\newif\ifwithappendix
\ifdefined\includeappendix
  \withappendixtrue
\else
  \withappendixfalse
\fi

\theoremstyle{plain}
\newtheorem{proposition}{Proposition}
\newtheorem{corollary}{Corollary}
\title{When Do AI Gains Become Broadly Shareable? A Policy Threshold for AI-Driven Automation}
\author{Aran Nayebi}
\affiliations{
Machine Learning Department and Neuroscience \& Robotics Institutes, \\
School of Computer Science, Carnegie Mellon University, Pittsburgh, PA, USA\\
\texttt{anayebi@cs.cmu.edu}
}

\begin{document}
\maketitle

\begin{abstract}
AI-driven automation generates broad-based social benefit only if technical gains become visible, durable, and publicly claimable. We develop a policy-facing stress test by extending a standard task-automation growth model with an AI capability parameter that raises productivity on automatable tasks while holding the set of tasks fixed. The exercise is intentionally limited: it is not a forecast of AI timelines or a full welfare analysis, but a way to identify which institutions determine whether AI rents can support broad transfers. Calibrated to U.S. quantities, the model shows that capability alone is not decisive. Public capture, deployment costs, automation scope, and market structure jointly determine when AI gains become shareable. The main policy lesson is that moving from low to moderate public capture (33\%) can substitute for substantial AI capability growth, while pushing capture further to full nationalization yields smaller gains, especially if deployment or safety costs are high. Competition policy also has distributional consequences: opening concentrated AI markets may improve fairness and resilience, but can reduce the rent pool unless alternative public-claim institutions are built. Cross-nationally, tax-heavy systems lower the needed AI capability threshold through stronger effective revenue collection, while Singaporean and Abu Dhabi-style public-asset models show that governments can also capture AI gains through ownership and investment returns rather than taxes alone. Our framework therefore identifies which levers governments can act on now to make future AI gains easier to measure, claim, and distribute broadly.
\end{abstract}

\section{Introduction}
AI governance debates often treat technical capability as the central variable: if AI becomes sufficiently powerful, then society can decide how to distribute the gains~\citep{dafoe2018ai,openai_charter,openai_agi_beyond,okeefe2020windfall,anderljung2023frontier}. 
This paper starts from the \emph{opposite} premise. 
Capability growth matters, but it does not by itself determine whether AI-driven productivity gains become broad public benefit, concentrated private rent, lower prices, worker augmentation, or offshored profit. 
That outcome depends on institutions: taxation, public ownership, anti-avoidance capacity, market concentration, procurement rules, and the design of AI deployment. 
Foundation models already sit inside cloud contracts, chip supply chains, enterprise software, labor processes, and public-facing services; the resulting rents are shaped by technical performance, but also by market concentration, ownership structure, and tax enforcement across the AI value chain \citep{CMA_FM_Update_2024,OECD_AI_Infrastructure_2025,OECD_AI_Downstream_2025}.

We develop a policy benchmark for this institutional problem. 
Rather than asking only how ``transformative'' AI systems may become, our benchmark asks: \textbf{Under what institutional conditions do AI-generated rents become sufficiently observable, capturable, and durable to support broad-based social benefit?} 
Throughout, ``AI rents'' denotes AI-attributable capital income available for public capture; Proposition~\ref{prop:oligopoly} separately adds pure economic profit created by market power.
Our benchmark links AI capability to the public capture of automation rents, but it is not tied to any single redistributive instrument. 
A broad transfer can be implemented as an AI dividend, a sovereign wealth dividend, a refundable tax credit, a negative income tax, or a universal basic income. 
Therefore, what we consider here is the \emph{size} of the broad-based transfer benchmark, and leave the precise administrative form used to distribute it to be flexibly chosen.

Specifically, our contribution is to extend traditional task-automation growth models with an AI capability parameter and use the resulting threshold condition to connect technical progress to institutional choices about public capture, market structure, and broad-based benefit. 
Task automation and capital deepening have long been central in macroeconomics \citep{solow1956contribution,solow1957technical,baumol1967macroeconomics,zeira1998workers,acemoglu2018race}. 
Recent AI-growth models and reviews ask how automation, new tasks, or explosive growth may change aggregate output \citep{aghion2017artificial,erdil2023explosive}. 
Our question is complementary: given a stream of AI-generated rents, we ask when existing institutions can convert those rents into broad social benefit. 
To make the benchmark demanding and interpretable, we calibrate it to the standardized 11\%-of-GDP UBI (Universal Basic Income)-sized transfer~\citep{UBICenterWhyNotUBI2019,BEA_GDP_Q3_2024}, while emphasizing that UBI is a yardstick rather than the paper's normative endpoint. 
To isolate the institutional problem, we hold the task frontier fixed and do not assume that AI creates new jobs, new tasks, or compensating labor demand. 
This makes the model a conservative policy \emph{stress test} rather than a speculative forecast.

Within that environment, we derive a \emph{closed-form} capability threshold linking AI productivity on automatable tasks to effective public capture, operating and compliance costs, automation intensity, and market structure. 
This reframing matters because these quantities correspond to \emph{observable policy levers}. 
Public capture depends on tax enforcement, withholding rules, public equity, licensing, and anti-avoidance capacity \citep{GAO-23-105384,OECD_Corporate_Tax_2025,OECD_WHT_2025,OECD_Global_Minimum_Tax_2026}. 
Operating costs depend on compute, inference, safety, energy, and compliance burdens. 
Market structure affects whether AI rents appear as competitive surplus, concentrated profit, or bargaining power within the AI value chain \citep{CMA_FM_Update_2024,OECD_AI_Infrastructure_2025,OECD_AI_Downstream_2025}. 
The benchmark therefore does not ask only whether AI becomes ``transformative''; rather, it asks whether governments can observe, capture, and govern the rents generated by AI deployment.

Using current U.S. macroeconomic quantities, our central calibration implies that supporting an 11\%-of-GDP broad transfer would require AI to reach roughly 5-7 times today's automation productivity on automatable tasks. Raising effective public capture from roughly 15\% to 33\% most significantly lowers the threshold to about $3.2\times$, with diminishing returns when AI companies are fully nationalized.
Market concentration lowers the funding threshold by expanding rents, but this is a fiscal result rather than a welfare endorsement. 
The same concentration that expands the capturable rent pool can also create exclusion, lock-in, dependence, and bottlenecks in AI infrastructure and downstream markets \citep{CMA_FM_Update_2024,OECD_AI_Infrastructure_2025,OECD_AI_Downstream_2025}. 
The policy implication is that capability policy, competition policy, and distribution policy must be designed \emph{together}.

We then use our benchmark comparatively, interpreting it through U.S., Nordic, Japanese, Singaporean, and Abu Dhabi-style institutional baselines. 
Recent comparative governance work emphasizes that generative-AI governance principles vary across regions, that compute-based governance capacity is unevenly distributed, and that systematic monitoring infrastructure remains underdeveloped \citep{LunaTXJ24,LehdonvirtaWH24,ArnoldSSLMSJLPE24}. 
Our benchmark complements that literature by translating institutional differences into capability thresholds for broad-based benefit. 
Countries with different levels of effective public capture, anti-avoidance capacity, public-asset ownership, and value-chain position do not face the same boundary between privately captured AI rents and publicly shareable AI gains.

The result is \emph{not} a prediction of when AI will fund a transfer program, nor a decision rule that mechanically determines policy. 
It is a diagnostic tool for asking which institutional constraints make broad-based AI benefit more or less fiscally plausible. 
Current empirical evidence suggests that many AI deployments still augment workers rather than replace them outright \citep{Brynjolfsson2023GenerativeAI,Dillon2025ShiftingWork,Anthropic_Economic_Index_2026}. 
Recent work from METR also suggests that governance-relevant capability measurement should be task-based and deployment-relevant, rather than limited to static model-evaluation scores \citep{METR_2025}. 
We therefore present our framework as a policy test for AI-driven automation, connecting fiscal feasibility to capability monitoring, anti-avoidance enforcement, public-asset strategies, procurement, competition policy, and labor augmentation.

\section{Observable Governance Levers}
Before turning to the formal model, Table~\ref{tab:measurement} maps our key parameters to observables and institutional settings. 
We place this map first because the model is meant to organize and \emph{prioritize} real policy levers, not reduce governance to a single capability score.

\begin{table}[t]
\centering
\small
\setlength{\tabcolsep}{4pt}
\begin{tabular}{p{0.12\linewidth} p{0.25\linewidth} p{0.51\linewidth}}
\hline
Lever & Meaning in our model & Observable proxy \\
\hline
$\gamma_t$ & AI productivity on automated tasks & Task-duration success rates, deployment benchmarks, and economically meaningful evaluations of autonomous task completion \citep{METR_2025}. \\
$\alphabar$ & Share of tasks that can be automated & Task and occupation exposure measures from workforce surveys, firm adoption data, and task-level automation studies \citep{WEF_FutureOfJobs_2023}. \\
$\Theta$ & Effective public capture of AI rents & Effective corporate tax collection, withholding regimes, public equity claims, and anti-avoidance enforcement outcomes \citep{GAO-23-105384,OECD_Corporate_Tax_2025,OECD_WHT_2025}. \\
$c$ & Operating and compliance cost share & Compute, inference, safety, energy, and regulatory expenditures as a share of gross AI revenue or rent. \\
$\psi_t$ & Augmentation before full automation & Worker-level productivity gains, time savings, and changes in task mix under AI assistance \citep{Brynjolfsson2023GenerativeAI,Dillon2025ShiftingWork,Anthropic_Economic_Index_2026}. \\
\hline
\end{tabular}
\caption{\textbf{Governance levers can be observed rather than treated as abstractions.} Each model parameter can be tied to a concrete measurement or institutional proxy.}
\label{tab:measurement}
\end{table}

Table~\ref{tab:measurement} also clarifies why we interpret AI as a value-chain and governance problem rather than a pure production-function problem. 
Capability gains do not automatically become public benefit. 
They may be dissipated by compute, safety, or compliance costs ($c$); retained privately when public capture ($\Theta$) is weak; or shifted across jurisdictions if tax and reporting rules are weak. 
Conversely, AI systems do not need to have automated the whole economy for governance questions to become pressing. 
Even partial automation ($\alphabar$) of high-value tasks can generate meaningful rents, while $\psi_t$ separately tracks gains from complementing workers before full automation \citep{CMA_FM_Update_2024,OECD_AI_Infrastructure_2025}.

For monitoring, $\gamma_t$ should be estimated sector by sector as quality-adjusted productivity on a fixed basket of automatable tasks relative to that sector's pre-AI automation baseline. An economy-wide index should use fixed value-added weights and report cross-sector dispersion; task-duration benchmarks are leading indicators rather than direct estimates of economic productivity. Effective public capture should be estimated as net AI-attributable taxes, withholding, royalties, licensing revenue, and public-asset returns, less AI-specific subsidies, divided by the corresponding AI-attributable capital-income base. The statutory rates used below are therefore scenario anchors, not estimates of national $\Theta$.

Seen this way, Table~\ref{tab:measurement} is also a map of institutional responsibility. Model developers and cloud providers influence $c$ through compute efficiency, safety practices, and compliance architecture. Downstream deployers influence $\alphabar$ and $\psi_t$ through workflow redesign, staffing choices, and adoption sequencing. Tax and finance authorities influence $\Theta$ through reporting, withholding, and anti-avoidance enforcement. The same deployment decision can therefore affect both the distribution of benefits and the amount of rent available to share in the first place \citep{CMA_FM_Update_2024,OECD_AI_Infrastructure_2025,OECD_AI_Downstream_2025}.

Finally, beginning with this table helps us keep two transition channels conceptually separate. 
The first is \emph{augmentation}, which we represent with $\psi_t$ and which is already visible in current deployments. 
The second is \emph{automation rent extraction}, which depends on both $\bar{\alpha}$, the share of tasks exposed to automation, and $\gamma_t$, the productivity of AI on those automated tasks.
The same solvency benchmark applies across both these channels, with augmentation shaping near-term output and automation rents determining the longer-run funding base.

\subsection{A Large UBI Benchmark as a Useful Common Yardstick}
A calibration needs a common fiscal scale. We use an 11\%-of-GDP UBI-sized transfer because it is demanding enough to be policy-relevant, legible across disciplines, and administratively neutral in the model. It is a yardstick rather than a recommendation: the same aggregate funding condition can support AI dividends, refundable credits, wage supplements, public services, or other broad-benefit instruments.

This separation also clarifies our normative scope. The threshold is a positive fiscal-feasibility result conditional on the premise that some AI gains should be shared broadly; it does not identify the welfare-optimal transfer, delivery mechanism, or recipients. Readers therefore need not endorse UBI to use the benchmark.

Equation~\eqref{eq:threshold} also makes smaller targets transparent. Holding other parameters fixed, a transfer share $b=B/Y_t$ rescales as
\begin{equation}
\gamma_t^\star(b)=\gamma_t^\star(0.11)
\left(\frac{b}{0.11}\right)^\sigma.
\label{eq:transfer-scaling}
\end{equation}
At the central $\sigma=0.66$, a 5\%-of-GDP target lowers the current-capture benchmark from $5.6\times$ to about $3.3\times$, and the one-third-capture benchmark from $3.2\times$ to about $1.9\times$. Thus targeted instruments are not merely qualitatively easier to fund; their thresholds can be read directly from the same model.

\section{Defining Our Conservative Automation Benchmark: Solow-Zeira Growth Model with an AI Capability Parameter}
\label{sec:benchmark}
\subsection{Holding the Task Frontier Fixed Gives a Policy Stress Test}
We model GDP $Y_t$ with a constant-elasticity-of-substitution (CES) production function, which aggregates task-level goods $X_{it}$ into total output. 
In policy terms, this lets us represent the economy as a collection of tasks that differ in whether output is produced by labor or by automated capital. 
Specifically, at date $t$ a unit-measure continuum of tasks $i \in [0,1]$ is aggregated through
\begin{equation}
Y_t = A_t \left(\int_0^1 X_{it}^{\rho} \, di \right)^{1/\rho},
\qquad
\rho = \frac{\sigma - 1}{\sigma} < 0,
\label{eq:ces}
\end{equation}
where $A_t = A_0 e^{gt}$ is Hicks-neutral progress with growth rate $g \in \mathbb R$ captures generic efficiency improvements (e.g. Moore's law, better chips, cheaper energy, learning curves) that raise output even when no new jobs are created.
$\sigma \in (0,1)$ is the elasticity of substitution across tasks, where a lower $\sigma$ means that tasks are harder to substitute for one another, so bottlenecks in non-automated work continue to matter even when automated tasks become much more productive. 
As in \citet{zeira1998workers}, each task is produced by either automated capital or labor:
\begin{equation}
X_{it} =
\begin{cases}
K_{it} & \text{if task $i$ is automated at time $t$},\\
L_{it} & \text{otherwise}.
\end{cases}
\label{eq:task-production}
\end{equation}
Here $K_{it}$ denotes capital services used for an automated task, including machines, software, compute, and AI-enabled systems, while $L_{it}$ denotes labor on a non-automated task. The production block is technology-agnostic: software-based generative AI, embodied robotics, and hybrid systems can all enter through deployment-specific estimates of $\gamma_t$ and $\alphabar$.

Specifically, we fix a fraction $\alphabar \in (0,1]$ of tasks as automatable. 
As a result, we do not need to assume full automation; rather, $\alphabar$ is itself a governance-relevant lever: it captures how much of the economy can, in principle, be shifted into the automated block. 
Holding it fixed gives a conservative benchmark, which allows us to ask whether broad-based transfers remain feasible even if AI automates only the currently automatable share and no new tasks emerge.

With uniform allocation across the automated and non-automated blocks, i.e. $K_{it} \equiv K_t / \alphabar$ for automated tasks and $L_{it} \equiv L$ for non-automated tasks, output simplifies to
\begin{equation}
Y_t = A_t \left(\alphabar^{1-\rho} K_t^{\rho} + (1-\alphabar)^{1-\rho} L^{\rho}\right)^{1/\rho}.
\label{eq:baseline-ces}
\end{equation}
Here $K_t$ denotes aggregate capital services allocated to the automated block, while $L$ denotes labor allocated to the non-automated block. 
The two terms inside the production function therefore correspond to the automated and non-automated parts of the economy. 
This simplified expression is useful because it makes the central governance tradeoff explicit: broad-based benefit depends not only on how capable AI becomes, but also on how much of the economy is exposed to automation and how much of the resulting capital income can be publicly captured.


\subsection{Capability Should Scale Automated-Task Productivity, Not Raw Capital}

We model AI capability $\gamma_t$ as \emph{raising the productivity of automated tasks}. 
The parameter $\gamma_t$ is measured relative to pre-AI automation, so $\gamma_t=1$ is the baseline and $\gamma_t>1$ means that AI makes already-automated tasks more productive. 
Rather than treating AI as simply adding more raw capital, we let capability scale the effective contribution of the automatable block:
\begin{equation*}
\alphabar^{1-\rho} K_t^{\rho}
\longmapsto
(\gamma_t\alphabar)^{1-\rho} K_t^{\rho}
=
\gamma_t^{1-\rho}\alphabar^{1-\rho}K_t^{\rho}.
\end{equation*}
The production function therefore becomes
\begin{equation}
Y_t = A_t \left(\gamma_t^{1-\rho}\alphabar^{1-\rho} K_t^{\rho} + (1-\alphabar)^{1-\rho} L^{\rho}\right)^{1/\rho},
\quad
\gamma_t \ge 1.
\label{eq:ai-ces}
\end{equation}

This specification has a direct policy interpretation:
AI capability does not expand the literal set of automatable tasks; that role is played by $\alphabar$; instead, $\gamma_t$ measures how much more economically important the automated block becomes once AI improves the productivity of tasks already exposed to automation. 
This distinction is useful because the governance question is not only how many tasks can be automated, but also how much rent those automated tasks can generate once AI systems become more capable.

The choice to scale the automated block rather than raw capital is also economically motivated. 
If we multiplied capital directly by $\gamma_t$, the capability boost would enter as $\gamma_t^{\rho}$ and would be mechanically attenuated when $\rho<0$. 
That would make higher AI capability look less important precisely in the case where tasks are hard to substitute for one another. 
For our purposes, that is the wrong object: we want $\gamma_t$ to capture the productivity of AI on automatable tasks, not simply the quantity of capital used in them.

The condition $\rho<0$, equivalently $\sigma<1$, means that tasks are gross complements. 
In policy terms, this captures the Baumol ``cost-disease'' effect~\citep{baumol1967macroeconomics}: even if AI makes automated tasks much cheaper or more productive, non-automated tasks such as care, oversight, judgment, service delivery, or other labor-intensive activities can remain economically important. 
As automated tasks become more productive, the economy does not necessarily become all automation all the way down. 
Instead, relative scarcity shifts toward the slower-moving non-automated block. 
This keeps the model from assuming that AI capability alone makes labor irrelevant, while still allowing higher $\gamma_t$ to expand the rent pool generated by automation.

Finally, we note that restricting attention to $\gamma_t\ge 1$ is without loss of generality. 
Values below one would mean that AI makes the automated block less productive than pre-AI automation. 
In the model, such a case can be absorbed by redefining the effective automatable share as $\alphabar'=\gamma_t\alphabar$, so it adds no distinct economics. 
We therefore treat $\gamma_t=1$ as the pre-AI automation baseline and $\gamma_t>1$ as any additional productivity boost delivered by AI, if any.


\subsection{Effective Public Capture is a Key Fiscal Lever}

To determine how much rent can be shared, we also need a simple way to summarize the economy's capital base. 
We assume that a constant share $s$ of output is saved and invested, while existing capital depreciates at rate $\delta$ via the standard Solow capital-accumulation difference equation~\citep{solow1956contribution,solow1957technical}:
\begin{equation}
K_{t+1} = sY_t + (1-\delta)K_t.
\label{eq:capital-law}
\end{equation}
This standard accumulation equation says that next period's capital equals new investment plus the undepreciated part of the current capital stock.

Once AI capability $\gamma_t$ stabilizes, or grows at any constant factor $G_\gamma := \gamma_{t+1}/\gamma_t > 0$ while the capital stock adjusts at $G_K := K_{t+1}/K_t = G_\gamma^{1/(1-\sigma)}$, so that $Y_{t+1}=e^{g}Y_t$ in either case, this linear difference equation forces the capital-output ratio to a Solow limit of
\begin{equation}
\frac{K_t}{Y_t} \xrightarrow{t \to \infty} \kappabar = \frac{s}{e^{g}-1+\delta} 
\;\overset{\lvert g\rvert\ll 1}{\approx}\; \frac{s}{g+\delta}.
\label{eq:kappa}
\end{equation}
We use this ratio to express the rent-generating automated capital base relative to the size of the economy. 
This keeps the threshold in terms of observable macro quantities rather than requiring a separate forecast of the total capital stock.

We model the public sector as capturing a share $\Theta \in (0,1]$ of AI rents, net of an operating and compliance cost share $c \in [0,1)$. 
We interpret $\Theta$ broadly as \emph{effective} public capture through taxation, withholding, anti-avoidance enforcement, public ownership, or equivalent claims on AI-generated rents. 
This broader interpretation is important: in the comparative section below, we use real-world tax regimes as stylized baselines for $\Theta$, but we also distinguish tax-heavy systems from public-asset models whose effective public claim can exceed what headline tax rates alone suggest.

\section{Identifying Tractable Governance Thresholds}
Our model, defined in \S\ref{sec:benchmark}, yields a transparent benchmark for the capability required to finance a transfer benchmark $B$:

\begin{proposition}[Capability threshold]
\label{prop:threshold}
In the economy described by \eqref{eq:ai-ces}--\eqref{eq:kappa}, a transfer benchmark $B$ is fiscally sustainable in period $t$ if and only if
\begin{equation}
\gamma_t \ge \gstar =
\left(
\frac{\bstar}
{\Theta (1-c)\alphabar^{1-\rho} A_t^{\rho} \kappabar^{\rho}}
\right)^{\sigma}, \quad \sigma=\frac{1}{1-\rho} > 0.
\label{eq:threshold}
\end{equation}
\end{proposition}
\textbf{Proofs of all propositions and corollaries are provided in the Technical Appendix.}

Proposition~\ref{prop:threshold} is the benchmark around which the rest of our analysis turns. 
It says that the AI capability threshold $\gstar$ is \emph{multiplicatively} shaped by public capture ($\Theta$), AI training and safety costs ($c$), automation intensity ($\alphabar$), and the capital-output ratio ($\kappabar$). 
In particular, AI capability and public capture are not substitutes in any simple political sense. 
A society can lower the required capability by raising capture, or lower the required capture by waiting for more capable AI, but the trade-off is dictated by \eqref{eq:threshold}.


\begin{corollary}[Comparative statics]
\label{cor:comparative}
For the threshold
\[
\gstar =
\left(
\frac{B/Y_t}
{\Theta(1-c)\alphabar^{1-\rho}A_t^\rho \kappabar^\rho}
\right)^\sigma,
\qquad
\kappabar = \frac{s}{e^g-1+\delta},
\]
the first-order effects are
\begin{align}
\frac{\partial \gstar}{\partial \Theta}
&=
-\frac{\sigma \gstar}{\Theta} < 0,
&
\frac{\partial \gstar}{\partial c}
&=
\frac{\sigma \gstar}{1-c} > 0,
\nonumber\\
\frac{\partial \gstar}{\partial \alphabar}
&=
-\frac{\gstar}{\alphabar} < 0,
&
\frac{\partial \gstar}{\partial \kappabar}
&=
-\frac{\sigma \rho \gstar}{\kappabar} > 0,
\label{eq:comparative}\\
\frac{\partial \gstar}{\partial s}
&=
-\frac{\sigma \rho \gstar}{s} > 0,
&
\frac{\partial \gstar}{\partial \delta}
&=
\frac{(\sigma-1)\gstar}{e^g-1+\delta} < 0.
\nonumber
\end{align}
\end{corollary}

The comparative statics give insight into governance: 
More public capture ($\Theta$) lowers the required AI capability for broad-based benefits. 
Higher operating or safety costs ($c$) raise it. 
Broader automation $\alphabar$ lowers it.
A higher capital-output ratio $\kappabar$ raises the threshold because the automated-capital rent share enters as $\kappabar^\rho$ with $\rho<0$: at a fixed capability level, additional capital reduces the marginal rent share of the automated block. 
The saving and depreciation effects follow through this same capital-output channel: higher saving $s$ raises $\kappabar$ and therefore raises the required capability, while higher depreciation $\delta$ lowers $\kappabar$ and therefore lowers it.

\begin{proposition}[Funding threshold under oligopoly]
\label{prop:oligopoly}
Suppose the AI-capital sector is supplied by Cournot firms with conduct parameter $\theta = \sum_i s_i^2 \in [0,1]$ and demand elasticity $\eps > 0$, so pure profit accounts for the share $\theta/\eps$ of output. If the public sector captures that pure-profit component in addition to the rent share in Proposition~\ref{prop:threshold}, then the transfer benchmark is sustainable if and only if
\begin{equation}
\gamma^{\star}_{\mathrm{oligo},t}
=
\left(
\frac{\bstar}{\Theta(1-c)}
\frac{1}{\alphabar^{1-\rho}A_t^{\rho}\kappabar^{\rho}}
-\frac{\theta/\eps}{\alphabar^{1-\rho}A_t^{\rho}\kappabar^{\rho}}
\right)^{\sigma}.
\label{eq:oligopoly}
\end{equation}
Whenever $\theta > 0$, $\gamma^{\star}_{\mathrm{oligo},t} < \gstar$.
\end{proposition}

Equation~\eqref{eq:oligopoly} should \emph{not} be read as saying that concentrated AI markets are good. 
It says only that, if a small number of firms earn large AI rents, there is a larger pool of money that governments could in principle capture and redistribute, thereby making a broad transfer easier to fund. 
But the same concentration can also make AI markets less fair and less resilient by increasing lock-in, limiting entry, raising switching costs, and making governments and users dependent on a few suppliers.

\begin{corollary}[Higher public capture lowers the necessary AI capability threshold]
\label{cor:theta}
If $\Theta_2 > \Theta_1$, then
\begin{equation}
\gamma_{2,t}^{\star} < \gamma_{1,t}^{\star}
\quad\text{and}\quad
\gamma_{1,t}^{\star} - \gamma_{2,t}^{\star}
=
C_t \left(\Theta_1^{-\sigma} - \Theta_2^{-\sigma}\right) > 0,
\label{eq:theta-cor}
\end{equation}
where
\begin{equation}
C_t =
\left(
\frac{\bstar}
{(1-c)\alphabar^{1-\rho}A_t^{\rho}\kappabar^{\rho}}
\right)^\sigma.
\end{equation}
\end{corollary}

Corollary~\ref{cor:theta} is especially useful for comparative analysis. It tells us that cross-national differences in effective public capture map directly into different capability thresholds. That is the compact formal statement behind the comparative discussion below: countries do not need the same level of AI capability to support broad-based benefit if their institutions turn rents into public revenue at different rates, as we now analyze below.

\section{Calibration and Sensitivity Analyses}
\label{sec:calibration}
\subsection{Moderate Public Capture Substantially Lowers the Needed Capability}
We start by calibrating the benchmark using current U.S. quantities where possible. 
The transfer benchmark is $\bstar \approx 0.11$, corresponding to a \$12{,}000 annual transfer for adults relative to 2024 U.S. GDP \citep{UBICenterWhyNotUBI2019,BEA_GDP_Q3_2024}. 
We set public capture $\Theta = 0.145$ using GAO evidence on effective federal corporate income taxation \citep{GAO-23-105384}. We use $\alphabar = 0.42$ based on automation-related task shares reported by the World Economic Forum \citep{WEF_FutureOfJobs_2023}. For substitution, we use the U.S. capital-labor elasticity range summarized by \citet{knoblach2020elasticity}, centered at $\sigma = 0.66$. We set $s = 0.22$ from U.S. gross capital formation \citep{WorldBank_GCF_US}, $g = 0.011$ from CBO productivity projections \citep{CBO_ProductivityForecast_2024}, and $\delta = 0.056$ from BEA fixed-asset data \citep{BEA_M1PTOTL1ES000,BEA_K1PTOTL1ES000}. Because frontier AI operating and compliance costs remain highly uncertain, we treat $c$ as a policy-sensitive range and use $c = 0.60$ as our central benchmark, conservatively informed by third-party estimates of OpenAI's gross margins and inference costs~\citep{SacraOpenAI2025}.

\begin{figure*}[t]
\centering
\includegraphics[width=0.99\linewidth]{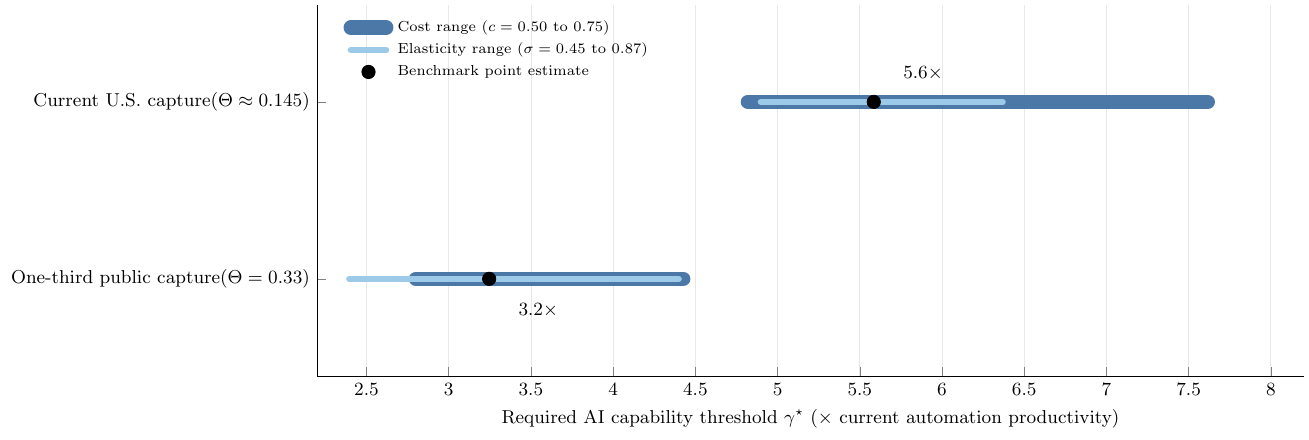}
\caption{\textbf{Moderate public capture can substitute for substantial capability growth.} Thick segments show the threshold range induced by operating-cost uncertainty, thin inner segments show the elasticity range, and dots show the benchmark point estimate. Under current U.S. capture, the required AI capability threshold is roughly $5.6\times$ today's automation productivity; under one-third public capture it falls to roughly $3.2\times$.}
\label{fig:benchmark}
\end{figure*}

That last choice is worth underscoring. We do not want the benchmark to smuggle in a false precision about operating costs. Compute, inference, safety, energy, evaluation, and compliance expenditures may all change quickly as AI diffuses into more sectors. In other words, $c$ is not just a nuisance parameter; it is one of the ways in which the institutional and technical shape of deployment enters the fiscal question. For that reason, we plot sensitivity bands in Figure~\ref{fig:benchmark} rather than a single point estimate.

Figure~\ref{fig:benchmark} summarizes the first-order benchmark. Under current public capture and benchmark costs, the threshold is about $5.6\times$ today's automation productivity on automatable tasks. Moving across plausible operating-cost values produces a range of roughly $4.8\times$ to $7.6\times$. Varying $\sigma$ across the \citet{knoblach2020elasticity} range yields a threshold between about $4.9\times$ and $6.4\times$. Raising public capture from about 15\% to one-third lowers the threshold to roughly $3.2\times$, with a broader sensitivity range of roughly $2.4\times$ to $4.4\times$ once elasticity and cost uncertainty are included. Institutional reform therefore changes the calibrated boundary by several multiples of today's automation productivity.

The policy significance of Figure~\ref{fig:benchmark} is not that it predicts a date on which AI will inevitably cross the line. We intentionally avoid that framing here. The point is instead that the relevant threshold is not obviously science-fictional in scale. In our benchmark, institutions move the threshold by roughly the same order of magnitude as many speculative arguments about capability growth. That is precisely why governance choices matter \emph{earlier} rather than later.

\subsection{Market Concentration Can Lower the Funding Threshold While Worsening Welfare}
\begin{figure*}[t]
\centering
\includegraphics[width=0.94\linewidth]{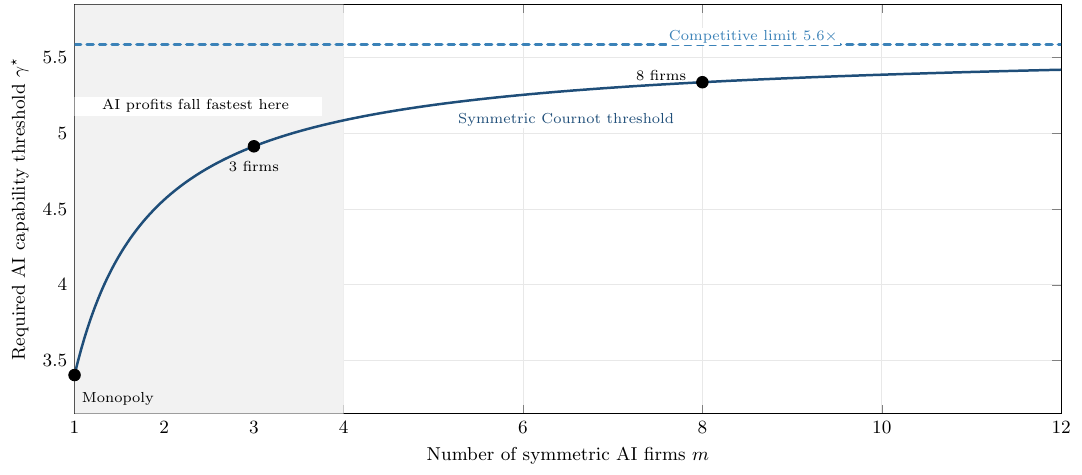}
\caption{\textbf{Competition policy and distribution policy have to be designed together.} The solid curve shows the symmetric-Cournot threshold when conduct scales as $\theta=1/m$ and the absolute demand elasticity is fixed at the \citet{ukaiimpact2023} midpoint of $|\eps|=1$; the dashed line marks the competitive benchmark. Moving from monopoly to a small oligopoly raises the required AI capability threshold sharply, while additional entry beyond that has smaller effects. The shaded region marks where competition reduces AI firms' excess profits most quickly.}
\label{fig:competition}
\end{figure*}

Figure~\ref{fig:competition} illustrates Proposition~\ref{prop:oligopoly}. Imposing the symmetric-Cournot benchmark $\theta=1/m$ makes the policy intuition transparent: most of the change happens when markets move from monopoly to a small oligopoly, not when they move from ten firms to twenty-five. The current AI sector sits much closer to concentrated oligopoly than to perfect competition in several layers of the stack, from specialized chips and cloud infrastructure to access pathways for frontier models \citep{CMA_FM_Update_2024,OECD_AI_Infrastructure_2025}. In our benchmark, that concentration lowers the \emph{funding} threshold because it creates larger profits. But it simultaneously raises governance concerns about dependency, foreclosure, and bargaining power within the AI value chain. Proposition~\ref{prop:oligopoly} establishes no welfare ranking: consumer surplus, innovation, worker power, resilience, and democratic accountability remain outside its fiscal comparison.

This is a genuinely double-edged result for AI governance. If regulators successfully open markets and reduce rents, then broad-based benefit does not become less desirable; it simply has to be financed through more robust public-capture instruments rather than by relying on concentrated markups. The right lesson is therefore not ``monopoly is good,'' but that competition governance and distributional governance cannot be analyzed in isolation.

\subsection{Moving from Low to Moderate Capture Matters More Than Chasing Full Nationalization}
\begin{figure*}[t]
\centering
\includegraphics[width=0.92\linewidth]{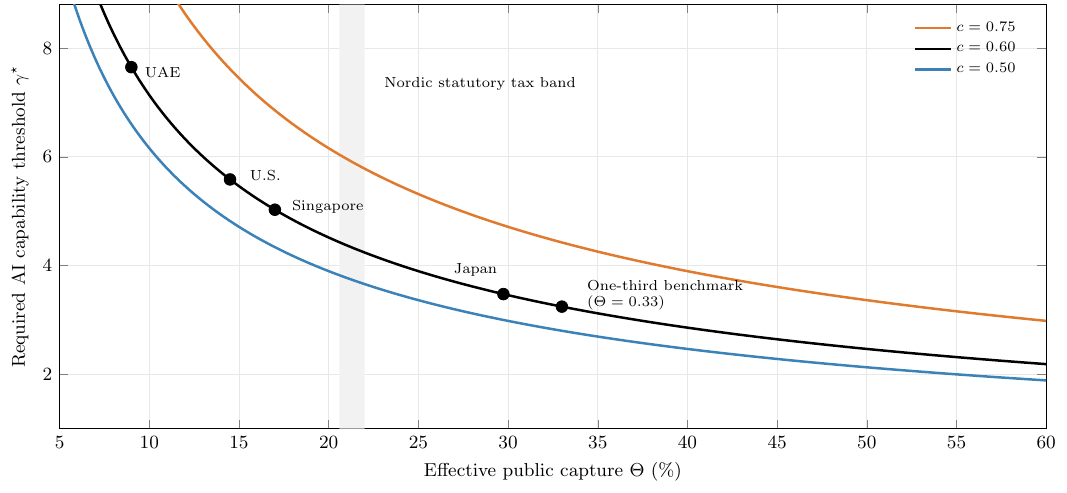}
\caption{\textbf{The biggest fiscal gains come from moving from low to moderate public capture.} The curves plot the required AI capability threshold as effective public capture rises under three operating-cost assumptions ($c=0.50,0.60,0.75$). Axis markers indicate stylized comparison points for the UAE, United States, Singapore, a Nordic tax band, Japan, and the one-third benchmark. Because $\Theta$ is broader than statutory taxation alone, Singaporean and Abu Dhabi-style public-asset institutions should be read as tax-only lower bounds rather than complete estimates of public claim on AI rents.}
\label{fig:ownership}
\end{figure*}

Figure~\ref{fig:ownership} shows that public capture is one of the strongest institutional levers in the model, but also that its gains are \emph{not} linear. Starting from the U.S. benchmark, raising effective capture lowers the threshold sharply at first and then more slowly. The one-third benchmark is therefore attractive not because it is magically optimal, but because it captures the steep part of the curve. Past that point, additional capture still helps, but the marginal reduction in $\gamma^\star$ is noticeably smaller, especially when $c$ is high.

That curvature is not incidental. Because $\gamma^\star = C_t \Theta^{-\sigma}$ and $\sigma \in (0,1)$, the largest threshold reductions occur when a country moves from very low capture toward moderate capture; later increments still help, but less dramatically. This is the formal reason the one-third benchmark matters so much in our calibration and why moving the threshold does not require full nationalization or anything close to it. Once operating and compliance costs are high, the practical challenge is not simply to raise headline rates, but to raise \emph{effective} capture while keeping administrative drag and avoidance manageable.

This is where Corollary~\ref{cor:theta} becomes useful for comparative policy analysis. 
Cross-national tax and ownership regimes provide a rough sense of how different states sit at different points on the curve, even though statutory tax rates do \emph{not} map one-to-one into $\Theta$, as it can depend on tax bases, deductions, enforcement, profit shifting, withholding rules, public ownership, and other claims on AI-generated value.
Sweden's official corporation tax rate is 20.6\% \citep{Sweden_Corporation_Tax_2025}, while Norway's business general-income rate is 22\% \citep{Norway_General_Income_2025}. Singapore's flat corporate income tax rate is 17\% \citep{Singapore_Corporate_Tax_2026}, and the UAE's federal corporate tax rate is 9\% above the relevant threshold \citep{UAE_Corporate_Tax_2026}. For Japan, JETRO reports a 29.74\% effective corporate tax rate for larger enterprises on a standard-rate basis \citep{JETRO_Japan_Corporate_Tax_2025}. Using those values only as stylized baselines for effective capture, the implied benchmark threshold is roughly $5.6\times$ under the current U.S.\ baseline, $5.0\times$ under Singapore's headline tax rate, $4.4\times$ under a Swedish-style rate, $4.2\times$ under a Norwegian-style rate, $3.5\times$ under the Japanese large-firm rate, $3.2\times$ at the one-third benchmark, and $7.7\times$ under the UAE's tax-only federal baseline.

\begin{table}[t]
\centering
\small
\begin{tabular}{p{0.42\linewidth} c c}
\hline
Illustrative baseline & $\Theta$ used & $\gamma^\star$ at $c=0.60$ \\
\hline
Current U.S. effective federal corporate tax & 0.145 & 5.6 \\
Singapore statutory corporate tax & 0.170 & 5.0 \\
Sweden statutory corporate tax & 0.206 & 4.4 \\
Norway business/general-income rate & 0.220 & 4.2 \\
Japan effective corporate tax rate (larger enterprises) & 0.297 & 3.5 \\
One-third public-capture benchmark $(\Theta = 0.33)$ & 0.330 & 3.2 \\
UAE federal corporate tax (tax-only baseline) & 0.090 & 7.7 \\
\hline
\end{tabular}
\caption{\textbf{Tax-heavy and public-asset states offer different routes to effective public capture.} These values use published tax rates as stylized reference points rather than direct estimates of effective AI-rent capture.}
\label{tab:comparative}
\end{table}

Table~\ref{tab:comparative} should therefore be read as a comparison of jurisdictional fiscal capacity: countries with stronger institutions for turning concentrated profits into public revenue need less AI capability to support the same transfer benchmark. It is not a claim that the state hosting or taxing an AI firm has the sole moral entitlement to globally produced value.

Singapore and Abu Dhabi sharpen that comparative reading because both illustrate public-asset pathways that a tax-only proxy understates. Singapore's Net Investment Returns Contribution allows the government to spend up to 50\% of expected long-run returns on reserves invested by GIC, MAS, and Temasek, which means public capture can occur through public assets as well as through ordinary corporate taxation \citep{Singapore_NIRC_2025,GIC_WhoWeAre_2026}. Abu Dhabi combines the UAE's low federal corporate tax with large sovereign investors such as ADIA, Mubadala, and ADQ, which are explicitly tasked with sustaining long-run prosperity, generating returns for the government, and diversifying the economy \citep{ADIA_Purpose_2026,Mubadala_About_2026,ADQ_WhoWeAre_2026}. The point is not that these institutions can be copied wholesale or that they already solve AI distribution. It is that low headline taxation and low effective public claim are not the same thing.

That distinction matters for current industrial-policy debates. Many proposals for AI dividends, public co-investment, sovereign or social wealth funds, data-center community-benefit agreements, and public-interest licensing are trying to move only three objects in our model: raise $\Theta$, keep $c$ from being shifted onto the public, or increase visible worker benefit during the augmentation phase. Our framework validates that instinct, but it also clarifies the constraint: these ideas matter only if they produce durable public claims on rent without adding so much leakage or administrative drag that the gains are dissipated.

The comparative lesson is also about sequencing. Jurisdictions starting near the U.S. baseline cannot count on capability growth alone if they want broad-based benefit in a politically relevant time frame; they need to move on capture and enforcement earlier. Jurisdictions already closer to the one-third benchmark ($\Theta = 0.33$) may get more from improving measurement, competition governance, and transition support, because additional gains from higher capture arrive more slowly once they are already on the flatter part of Figure~\ref{fig:ownership}. The same global technology trend therefore implies different immediate priorities across states.

That observation matters especially because AI rents are likely to be internationally mobile. Cross-border payments, licensing arrangements, cloud contracts, and intellectual-property location choices all affect where rents appear to arise and where they are taxed. OECD evidence on withholding taxes, treaty networks, and minimum-tax coordination exists precisely because firms can shift profits when public capture is fragmented or poorly coordinated \citep{OECD_Corporate_Tax_2025,OECD_WHT_2025,OECD_Global_Minimum_Tax_2026}. In an AI context, this means that public-capture policy is partly a domestic question and partly a co-ordination question. If countries compete to attract compute clusters or model firms by offering weak capture, the effective $\Theta$ relevant for broad-based benefit can stay low even as aggregate rents rise.

Comparative perspective also means recognizing that jurisdictions do not occupy the same place in the AI value chain. A country that hosts model labs, cloud regions, or valuable downstream platforms can often capture profits more easily than a country that mainly imports AI services. For the latter, the central issue is often less which headline rate to choose than whether source-based withholding, reporting obligations, and co-ordinated audit rules prevent value generated in domestic deployment from being booked elsewhere. Cross-national inequality therefore enters our benchmark through institutions as much as through capability: the same $\gamma_t$ can produce very different social outcomes depending on whether a state has the capacity to locate, measure, and claim part of the profit stream \citep{OECD_WHT_2025,OECD_Global_Minimum_Tax_2026}.

Recent comparative governance scholarship reaches a similar conclusion. Luna et al.\ show that governance processes and principles for generative AI already differ meaningfully across regions, while Lehdonvirta, Wu, and Hawkins show that the geography of AI compute creates unequal opportunities for states to govern AI systems through territorial control of infrastructure \citep{LunaTXJ24,LehdonvirtaWH24}. Our comparative thresholds translate those institutional asymmetries into a more concrete question: how much capability must a country wait for if it cannot already locate, tax, or own a meaningful share of AI rents?

Real-world public-rent institutions also illustrate why the model's collective-benefit channel is not merely hypothetical. Norway's Government Pension Fund Global is financed by state petroleum revenues and invests those proceeds for long-run public benefit \citep{Norway_Government_Pension_Fund_2026}. Petroleum rents are obviously not AI rents, and we do not mean to collapse the analogy. But the institutional lesson is important: concentrated economic rents can be converted into broad public assets through durable state capacity rather than through ad hoc redistribution alone. When public-asset pathways are used, however, governance quality matters too. The Santiago Principles exist precisely because sovereign wealth funds need transparent mandates, accountability, and risk-management discipline if they are to convert concentrated rents into legitimate public benefit \citep{IFSWF_Santiago_Principles_2026}. That helps explain why our model treats UBI as one benchmark inside a broader family of dispersion mechanisms rather than as the only available endpoint.

The practical implication is that distribution mechanisms should be matched to institutional maturity. Some countries may be better placed to rely on general taxation and existing transfer systems; others may find royalty-style mechanisms, sectoral levies, or publicly governed funds easier to administer. Our benchmark does not adjudicate among those designs. It clarifies that they are all attempts to move the same underlying lever: effective public capture of AI rents.

\section{Practical Policy Takeaways}
Our benchmark is most useful as a policy checklist. Broad-based benefit depends not only on how capable AI systems become, but also on whether governments can capture part of the resulting rent, keep deployment costs visible, preserve contestability where it matters, and make near-term augmentation beneficial enough to sustain legitimacy. These are not separate debates: they map onto public capture ($\Theta$), AI safety and operating costs ($c$), market concentration ($\theta$), demand elasticity ($\eps$), the share of tasks automated ($\alphabar$), and labor augmentation ($\psi_t$), and they therefore shift the capability threshold itself. The practical question is which of these levers a government can move before AI rents, contracts, and deployment patterns become institutionally entrenched.

\subsection{Make AI Rents Claimable Early}
The strongest direct lever is effective public capture, $\Theta$. The practical target is not a single tax instrument or maximal public ownership, but a credible institutional bundle that turns AI rents into durable public claims. Corporate taxation can do part of this work, but only if it is paired with anti-avoidance, withholding, disclosure, and minimum-tax coordination that limit leakage through royalties, transfer pricing, cloud contracts, and intellectual-property location choices \citep{OECD_Corporate_Tax_2025,OECD_WHT_2025,OECD_Global_Minimum_Tax_2026}. In our model, weak enforcement simply lowers effective $\Theta$ and raises the required $\gamma^\star$.

This implies a concrete administrative agenda. Tax authorities need enough reporting to distinguish domestic AI deployment revenue, cloud and inference payments, model-access royalties, and capital income booked through related entities. Public agencies can support that task by requiring large AI vendors to disclose contracting structures, compute-region dependencies, and material subcontractors when they sell into public or regulated sectors. The point is to make the AI value chain legible enough that ordinary fiscal tools can operate before rents become hard to locate.

Public-asset channels are the second route. Norway's petroleum fund, Singapore's public-investment model, and Abu Dhabi's sovereign investment institutions are not AI policies, but they show that concentrated rents can be converted into public balance-sheet claims when the state has durable investment and governance capacity \citep{Norway_Government_Pension_Fund_2026,Singapore_NIRC_2025,GIC_WhoWeAre_2026,ADIA_Purpose_2026,Mubadala_About_2026,ADQ_WhoWeAre_2026}. An AI analogue could take the form of royalties on scarce compute access, public equity in infrastructure, licensing fees, or a social wealth fund capitalized by AI-related revenues.

This is why the result does not imply full-scale nationalization. Moderate, well-governed public claims can move $\Theta$ without owning AI firms outright; poorly governed subsidies can raise $c$ or leave $\Theta$ effectively unchanged. The key design problem is to make public claims predictable enough for investment, but strong enough that profits from socially consequential automation are not locked into private balance sheets before institutions catch up.

The distinction between statutory and effective capture is crucial for innovation incentives. A levy that deters beneficial investment, induces relocation or avoidance, or is offset by subsidies may raise a headline rate without raising realized $\Theta$. Our model therefore does not identify an optimal tax rate. It shows what a realized public claim would do to fiscal feasibility; policy design must separately test investment, entry, location, and tax-base responses.

\subsection{Protect Public Benefit When Making Markets More Competitive}
Competition policy and distribution policy need to be designed together. Our oligopoly result is a funding result, not a welfare endorsement: concentration can lower the transfer-funding threshold by preserving rents, while also increasing dependence, exclusion, and lock-in. Competition authorities are therefore right to worry about AI infrastructure and downstream market power \citep{CMA_FM_Update_2024,OECD_AI_Infrastructure_2025,OECD_AI_Downstream_2025}. The policy question is what replaces concentrated rents if markets become more contestable.

One answer is stronger public capture. Another is interoperability policy that keeps diffusion broad while still preserving a public claim on the surplus. Public compute investment, open standards, model-access conditions, and public-interest licensing can reduce dependence on a few suppliers without giving up the distributional question. Antitrust success should therefore be measured not only by lower prices or more entrants, but also by whether the resulting market structure leaves institutions capable of financing broad benefit.

This is where the model disciplines a common policy instinct. ``More competition'' and ``more public revenue'' can pull in different directions if competition compresses rents faster than public institutions learn to capture them. A government that weakens gatekeeper power should simultaneously strengthen reporting, interoperability, and public-benefit claims, so diffusion gains are not purchased by making the future rent stream invisible or unclaimable.

\subsection{Write Public AI Contracts for Public Benefit}
Governments should treat public AI contracting as distribution policy, not only as technology purchasing. Public agencies, schools, courts, health systems, and benefits offices increasingly adopt AI through cloud contracts, model APIs, and application-layer vendors. Those contracts can lower effective deployment costs when they require interoperability, auditability, contestable switching, fallback procedures, and clear responsibility for errors. They can also raise $c$ when they create lock-in, opaque compliance burdens, or off-budget infrastructure costs \citep{CMA_FM_Update_2024,OECD_AI_Downstream_2025}.

Public value has to be visible before a large automation-profit pool exists. Better services, shorter waits, and meaningful error redress can build legitimacy for later public capture; surveillance, brittle automation, and vendor dependence can produce distrust. Procurement is therefore an early opportunity to make AI gains legible as public value rather than private cost cutting.

Public contracts should preserve switching rights, require audit logs for consequential uses, allocate responsibility for errors, limit exclusive data access, and disclose local energy or water burdens. These clauses shape $c$ and the legitimacy of $\Theta$.

\subsection{Make Augmentation Benefit Workers}
The near-term labor problem is likely to be augmentation before full replacement. Evidence from customer support, knowledge work, and observed model usage suggests that many current systems raise worker productivity or change task mix before they eliminate entire jobs \citep{Brynjolfsson2023GenerativeAI,Dillon2025ShiftingWork,Anthropic_Economic_Index_2026}. That is why $\psi_t$ matters. In the non-automated block, near-term labor augmentation can be represented as
\begin{equation}
(1-\alphabar)^{1-\rho}L^\rho
\longmapsto
(1-\alphabar)^{1-\rho}(\psi_t L)^\rho .
\label{eq:psi}
\end{equation}
This does not replace the automation-rent threshold based on $\gamma_t$. It adds a transition channel: AI can raise output and worker productivity before deeper substitution produces the dominant rent pool.

Policy should track wages, task discretion, service quality, and who captures time savings; whether output targets outpace pay; whether training reaches lower-wage workers; and whether workers can contest algorithmic management. Worker voice, sectoral training, support for smaller firms, and public-sector standards are therefore part of AI governance. A positive aggregate $\psi_t$ is not enough if job quality or bargaining power deteriorates. Steering $\psi_t$ toward workers can build legitimacy while institutions prepare to share later automation rents.

This also clarifies the conservative benchmark. Holding the task frontier fixed is not a forecast that new jobs will fail to emerge; it is a stress test for policy if job creation is not enough to absorb displacement.

\subsection{Track the Policy Levers Together}
The final takeaway is measurement. A capability threshold cannot be governed with a single benchmark score or countdown date. Policymakers need a dashboard linking task-based capability measures to sectoral deployment, profit location, market concentration, public receipts, worker outcomes, and infrastructure costs. METR's task-duration framing is closer to the model's $\gamma_t$ than decontextualized test scores, but it still needs institutional data about where economic value accrues \citep{METR_2025,ArnoldSSLMSJLPE24}.

Responsibility should be explicit. Statistical agencies and independent evaluation bodies should jointly estimate $\gamma_t$ and $\alphabar$; tax and finance authorities should publish effective $\Theta$; competition authorities should monitor $\theta$ and $\eps$; labor agencies should measure $\psi_t$; and infrastructure and sector regulators should report $c$. Each series should include methods, uncertainty intervals, and sectoral dispersion rather than one falsely precise national score.

The dashboard should trigger different responses depending on what moves. Rising capability with weak capture calls for disclosure and enforcement; concentrated profits with declining competition call for interoperability and market governance; visible augmentation with modest profits calls for training, bargaining, and public-service deployment standards. The point is not to find one magic number, but to know whether capability, capture, cost, market structure, or worker benefit is the binding constraint. Regular publication would make the benchmark useful before the threshold is crossed by revealing the binding institutional constraint.


\section{Limitations and Policy Directions}
Our benchmark is intentionally aggregate. We omit household incidence, endogenous task creation, and strategic profit shifting, though our supplementary Ramsey-Cass-Koopmans analysis (\S\ref{ss:endo}) shows lower \emph{endogenous} saving relaxes the AI feasibility constraint. These omissions matter for welfare and implementation; our work identifies the \emph{minimal} institutional variables richer models \emph{must} measure.

Key extensions sharpen instrument choice without sacrificing tractability: estimate sector-specific $\gamma_t$ from observed tasks \citep{METR_2025}; map $\Theta$ to taxes, withholding, royalties, and public assets; place energy, infrastructure, and safety spending inside $c$; and model the shift from augmentation $\psi_t$ to substitution. These extensions can turn the dashboard into policy triggers while preserving the result: governance choices materially shift when AI gains become broadly shareable.

\section*{Acknowledgements}
We thank Leo Kozachkov and Eric Young for feedback and acknowledge support from the Burroughs Wellcome Fund (CASI award) and Foresight Institute.

\clearpage
\section{Ethical Considerations and Distributional Scope}
Our fiscal threshold is conditional on a normative premise: some gains from socially consequential automation should be made broadly shareable. The equations do not establish that premise, select a welfare-optimal transfer, or determine who should receive it. UBI is a demanding common yardstick, not a claim that cash transfers always dominate public services, worker ownership, targeted credits, reparative investment, or other uses of public revenue. Democratic institutions still have to justify the objective, instrument, recipients, and tradeoffs.

The U.S. calibration likewise does not imply that the country hosting or taxing a leading AI firm has the sole legitimate claim on its profits. AI value may depend on workers, users, data and content contributors, supply chains, public research, knowledge commons, compute and energy infrastructure, natural resources, and affected communities across many jurisdictions. A nationally collected dollar is therefore a measure of fiscal capacity, not by itself a complete account of distributive justice. Treating globally produced value as belonging only to headquarters countries could deepen existing inequalities in compute access and governance capacity \citep{LehdonvirtaWH24,LunaTXJ24}.

Possible responses include source-based withholding, formulary allocation, royalties, community-benefit agreements, treaty coordination, and internationally governed funds. Our model does not privilege one arrangement. Under any of them, $\Theta$ should be interpreted as the share public institutions can direct after legitimate cross-border claims and transfers are recognized. Effective governance also requires representation and accountability for communities supplying labor, data, infrastructure, or resources, rather than assuming that efficient collection alone makes distribution fair.

Finally, neither of the model's two transition channels is ethically self-validating. The lower fiscal threshold under concentration is not an endorsement of monopoly: dependence, exclusion, surveillance, weak contestability, and political power may outweigh the fiscal advantage. Likewise, a positive augmentation effect $\psi_t$ can coexist with work intensification, lower autonomy, or weaker bargaining power. Public reporting, worker voice, contestability, and accessible mechanisms for redress are therefore safeguards for the same transition the benchmark describes, not optional additions after the threshold is crossed.
\clearpage
\bibliography{aaai2026}

@article{solow1956contribution,
  title={A contribution to the theory of economic growth},
  author={Solow, Robert M},
  journal={The quarterly journal of economics},
  volume={70},
  number={1},
  pages={65--94},
  year={1956},
  publisher={MIT press}
}

@article{solow1957technical,
  title={Technical change and the aggregate production function},
  author={Solow, Robert M},
  journal={The review of Economics and Statistics},
  volume={39},
  number={3},
  pages={312--320},
  year={1957},
  publisher={JSTOR}
}

@article{baumol1967macroeconomics,
  title={Macroeconomics of unbalanced growth: the anatomy of urban crisis},
  author={Baumol, William J},
  journal={The American economic review},
  volume={57},
  number={3},
  pages={415--426},
  year={1967},
  publisher={JSTOR}
}

@article{zeira1998workers,
  title={Workers, machines, and economic growth},
  author={Zeira, Joseph},
  journal={The Quarterly Journal of Economics},
  volume={113},
  number={4},
  pages={1091--1117},
  year={1998},
  publisher={MIT Press}
}

@article{erdil2023explosive,
  title={Explosive growth from AI automation: A review of the arguments},
  author={Erdil, Ege and Besiroglu, Tamay},
  journal={arXiv preprint arXiv:2309.11690},
  year={2023}
}

@misc{UBICenterWhyNotUBI2019,
  title  = {Why Not UBI?},
  author = {{UBI Center}},
  year   = {2019},
  month  = {9},
  url    = {https://www.ubicenter.org/what-is-ubi/why-not-ubi/}
}

@misc{BEA_GDP_Q3_2024,
  author       = {{U.S. Bureau of Economic Analysis}},
  title        = {Gross Domestic Product, Third Quarter 2024 (Advance Estimate)},
  year         = {2024},
  month        = {10},
  url          = {https://www.bea.gov/news/2024/gross-domestic-product-third-quarter-2024-advance-estimate}
}

@techreport{GAO-23-105384,
  author      = {{U.S. Government Accountability Office}},
  title       = {Corporate Income Tax: Effective Rates Before and After 2017 Law Change},
  institution = {U.S. Government Accountability Office},
  year        = {2022},
  month       = {12},
  number      = {GAO-23-105384},
  url         = {https://www.gao.gov/products/gao-23-105384}
}

@misc{SacraOpenAI2025,
  author       = {{Sacra}},
  title        = {OpenAI: Revenue, Valuation \& Growth Rate},
  year         = {2025},
  url          = {https://sacra.com/c/openai/}
}

@techreport{WEF_FutureOfJobs_2023,
  title        = {The Future of Jobs Report 2023},
  author       = {{World Economic Forum}},
  institution  = {World Economic Forum},
  year         = {2023},
  month        = {4},
  url          = {https://www.weforum.org/publications/the-future-of-jobs-report-2023/}
}

@article{knoblach2020elasticity,
  title={The elasticity of substitution between capital and labour in the US economy: A meta-regression analysis},
  author={Knoblach, Michael and Roessler, Martin and Zwerschke, Patrick},
  journal={Oxford Bulletin of Economics and Statistics},
  volume={82},
  number={1},
  pages={62--82},
  year={2020},
  publisher={Wiley Online Library}
}

@misc{WorldBank_GCF_US,
  author       = {{World Bank}},
  title        = {Gross Capital Formation (\% of GDP) - United States},
  year         = {2025},
  url          = {https://data.worldbank.org/indicator/NE.GDI.TOTL.ZS?locations=US}
}

@techreport{Brynjolfsson2023GenerativeAI,
  author       = {Erik Brynjolfsson and Danielle Li and Lindsey R. Raymond},
  title        = {Generative AI at Work},
  institution  = {National Bureau of Economic Research},
  year         = {2023},
  month        = {4},
  number       = {31161},
  doi          = {10.3386/w31161},
  url          = {https://www.nber.org/papers/w31161}
}

@techreport{CBO_ProductivityForecast_2024,
  author       = {{Congressional Budget Office}},
  title        = {CBO's Economic Forecast: Understanding Productivity Growth},
  institution  = {Congressional Budget Office},
  year         = {2024},
  month        = {7},
  number       = {60515},
  url          = {https://www.cbo.gov/publication/60515}
}

@misc{BEA_M1PTOTL1ES000,
  author       = {{U.S. Bureau of Economic Analysis}},
  title        = {Current-Cost Depreciation of Fixed Assets: Private},
  year         = {2024},
  month        = {10},
  url          = {https://fred.stlouisfed.org/series/M1PTOTL1ES000}
}

@misc{BEA_K1PTOTL1ES000,
  author       = {{U.S. Bureau of Economic Analysis}},
  title        = {Current-Cost Net Stock of Fixed Assets: Private},
  year         = {2024},
  month        = {10},
  url          = {https://fred.stlouisfed.org/series/K1PTOTL1ES000}
}

@misc{BLS_MFPNFBS_2025,
  author       = {{U.S. Bureau of Labor Statistics}},
  title        = {Private Nonfarm Business Sector: Total Factor Productivity},
  year         = {2025},
  month        = {3},
  url          = {https://fred.stlouisfed.org/series/MFPNFBS}
}

@misc{CMA_FM_Update_2024,
  author       = {{Competition and Markets Authority}},
  title        = {AI Foundation Models: Update Paper},
  year         = {2024},
  month        = {4},
  url          = {https://www.gov.uk/government/publications/ai-foundation-models-update-paper}
}

@techreport{OECD_AI_Infrastructure_2025,
  author       = {{Organisation for Economic Co-operation and Development}},
  title        = {Competition in Artificial Intelligence Infrastructure},
  institution  = {OECD},
  year         = {2025},
  month        = {11},
  number       = {OECD Roundtables on Competition Policy Papers No. 330},
  doi          = {10.1787/623d1874-en},
  url          = {https://www.oecd.org/en/publications/competition-in-artificial-intelligence-infrastructure_623d1874-en.html}
}

@techreport{OECD_AI_Downstream_2025,
  author       = {{Organisation for Economic Co-operation and Development}},
  title        = {Artificial Intelligence and Competitive Dynamics in Downstream Markets},
  institution  = {OECD},
  year         = {2025},
  month        = {11},
  number       = {OECD Roundtables on Competition Policy Papers No. 331},
  doi          = {10.1787/ccf0624a-en},
  url          = {https://www.oecd.org/en/publications/artificial-intelligence-and-competitive-dynamics-in-downstream-markets_ccf0624a-en.html}
}

@misc{METR_2025,
  author       = {{Model Evaluation and Threat Research}},
  title        = {Measuring {AI} Ability to Complete Long Tasks},
  year         = {2025},
  month        = {3},
  url          = {https://metr.org/blog/2025-03-19-measuring-ai-ability-to-complete-long-tasks/}
}

@misc{Anthropic_Economic_Index_2026,
  author       = {{Anthropic}},
  title        = {Anthropic Economic Index Report: Learning Curves},
  year         = {2026},
  month        = {3},
  url          = {https://www.anthropic.com/research/economic-index-march-2026-report}
}

@techreport{Dillon2025ShiftingWork,
  author       = {Eleanor W. Dillon and Sonia Jaffe and Nicole Immorlica and Christopher T. Stanton},
  title        = {Shifting Work Patterns with Generative {AI}},
  institution  = {National Bureau of Economic Research},
  year         = {2025},
  month        = {5},
  number       = {33795},
  doi          = {10.3386/w33795},
  url          = {https://www.nber.org/papers/w33795}
}

@techreport{OECD_Corporate_Tax_2025,
  author       = {{Organisation for Economic Co-operation and Development}},
  title        = {Corporate Tax Statistics 2025},
  institution  = {OECD},
  year         = {2025},
  month        = {11},
  doi          = {10.1787/6a915941-en},
  url          = {https://www.oecd.org/en/publications/corporate-tax-statistics-2025_6a915941-en.html}
}

@misc{OECD_WHT_2025,
  author       = {{Organisation for Economic Co-operation and Development}},
  title        = {Withholding Tax Rates and Tax Treaties: Corporate Tax Statistics 2025},
  year         = {2025},
  month        = {11},
  url          = {https://www.oecd.org/en/publications/corporate-tax-statistics-2025_6a915941-en/full-report/withholding-tax-rates-and-tax-treaties_e2216eab.html}
}

@misc{OECD_Global_Minimum_Tax_2026,
  author       = {{Organisation for Economic Co-operation and Development}},
  title        = {International Community Agrees Way Forward on Global Minimum Tax Package},
  year         = {2026},
  month        = {1},
  url          = {https://www.oecd.org/en/about/news/press-releases/2025/12/international-community-agrees-way-forward-on-global-minimum-tax-package.html}
}

@misc{Sweden_Corporation_Tax_2025,
  author       = {{Swedish Agency for Economic and Regional Growth}},
  title        = {Taxes and Contributions for Limited Companies},
  year         = {2025},
  month        = {10},
  url          = {https://verksamt.se/en/taxes-contributions/limited-companies}
}

@misc{Norway_General_Income_2025,
  author       = {{The Norwegian Tax Administration}},
  title        = {General Income},
  year         = {2025},
  url          = {https://www.skatteetaten.no/en/rates/general-income/}
}

@misc{Singapore_Corporate_Tax_2026,
  author       = {{Inland Revenue Authority of Singapore}},
  title        = {Corporate Income Tax Rates},
  year         = {2026},
  month        = {4},
  url          = {https://www.iras.gov.sg/quick-links/tax-rates/corporate-income-tax-rates}
}

@misc{Singapore_NIRC_2025,
  author       = {{Ministry of Finance Singapore}},
  title        = {Net Investment Returns Contribution},
  year         = {2025},
  month        = {10},
  url          = {https://www.mof.gov.sg/policies/reserves/net-investment-returns-contribution/}
}

@misc{GIC_WhoWeAre_2026,
  author       = {{GIC Private Limited}},
  title        = {Who We Are},
  year         = {2026},
  url          = {https://www.gic.com.sg/who-we-are/}
}

@misc{JETRO_Japan_Corporate_Tax_2025,
  author       = {{Japan External Trade Organization}},
  title        = {Overview of Corporate Income Taxes (Corporate Tax, Corporate Inhabitant Tax, Enterprise Tax)},
  year         = {2025},
  url          = {https://www.jetro.go.jp/en/invest/setting_up/section3/page3.html}
}

@misc{Norway_Government_Pension_Fund_2026,
  author       = {{Norwegian Ministry of Finance}},
  title        = {The Government Pension Fund (Statens pensjonsfond)},
  year         = {2026},
  url          = {https://www.regjeringen.no/en/dep/fin/about-the-ministry/subordinateagencies/the-government-pension-fund-/id270410/}
}

@misc{UAE_Corporate_Tax_2026,
  author       = {{Ministry of Finance - United Arab Emirates}},
  title        = {Corporate Tax in the UAE},
  year         = {2026},
  month        = {4},
  url          = {https://mof.gov.ae/corporate-tax/}
}

@misc{ADIA_Purpose_2026,
  author       = {{Abu Dhabi Investment Authority}},
  title        = {Purpose},
  year         = {2026},
  url          = {https://www.adia.ae/en/purpose}
}

@misc{Mubadala_About_2026,
  author       = {{Mubadala Investment Company}},
  title        = {About Mubadala},
  year         = {2026},
  url          = {https://www.mubadala.com/who-we-are/about-mubadala}
}

@misc{ADQ_WhoWeAre_2026,
  author       = {{ADQ}},
  title        = {Who We Are},
  year         = {2026},
  url          = {https://www.adq.ae/about-adq/who-we-are/}
}

@misc{IFSWF_Santiago_Principles_2026,
  author       = {{International Forum of Sovereign Wealth Funds}},
  title        = {Santiago Principles},
  year         = {2026},
  url          = {https://www.ifswf.org/santiago-principles}
}

@article{ArnoldSSLMSJLPE24,
  author       = {Zachary Arnold and Daniel S. Schiff and Kaylyn Jackson Schiff and Brian Love and Jennifer Melot and Neha Singh and Lindsay Jenkins and Ashley Lin and Konstantin Pilz and Ogadinma Enweareazu and Tyler Girard},
  title        = {Introducing the {AI} Governance and Regulatory Archive ({AGORA}): An Analytic Infrastructure for Navigating the Emerging {AI} Governance Landscape},
  journal      = {Proceedings of the AAAI/ACM Conference on AI Ethics and Society},
  year         = {2024},
  volume       = {7},
  number       = {1},
  pages        = {39--48},
  month        = {10},
  publisher    = {Association for the Advancement of Artificial Intelligence},
  doi          = {10.1609/aies.v7i1.31615},
  url          = {https://ojs.aaai.org/index.php/AIES/article/view/31615}
}

@article{LehdonvirtaWH24,
  author       = {Vili Lehdonvirta and Boxi Wu and Zoe Hawkins},
  title        = {Compute North vs. Compute South: The Uneven Possibilities of Compute-based {AI} Governance Around the Globe},
  journal      = {Proceedings of the AAAI/ACM Conference on AI Ethics and Society},
  year         = {2024},
  volume       = {7},
  number       = {1},
  pages        = {828--838},
  month        = {10},
  publisher    = {Association for the Advancement of Artificial Intelligence},
  doi          = {10.1609/aies.v7i1.31683},
  url          = {https://ojs.aaai.org/index.php/AIES/article/view/31683}
}

@article{LunaTXJ24,
  author       = {Jose Luna and Ivan Tan and Xiaofei Xie and Lingxiao Jiang},
  title        = {Navigating Governance Paradigms: A Cross-Regional Comparative Study of Generative {AI} Governance Processes \& Principles},
  journal      = {Proceedings of the AAAI/ACM Conference on AI Ethics and Society},
  year         = {2024},
  volume       = {7},
  number       = {1},
  pages        = {917--931},
  month        = {10},
  publisher    = {Association for the Advancement of Artificial Intelligence},
  doi          = {10.1609/aies.v7i1.31692},
  url          = {https://ojs.aaai.org/index.php/AIES/article/view/31692}
}

@article{xepapadeas2016spatial,
  title={Spatial growth with exogenous saving rates},
  author={Xepapadeas, Anastasios and Yannacopoulos, AN},
  journal={Journal of Mathematical Economics},
  volume={67},
  pages={125--137},
  year={2016},
  publisher={Elsevier}
}

@techreport{ukaiimpact2023,
  title        = {UK Artificial Intelligence Regulation Impact Assessment},
  author  = {{UK Department for Science, Innovation and Technology}},
  institution  = {UK Department for Science, Innovation and Technology},
  year         = {2023},
  month        = {3},
  url          = {https://assets.publishing.service.gov.uk/media/6424208f3d885d000cdadddf/uk_ai_regulation_impact_assessment.pdf}
}

@article{acemoglu2018race,
  title={The race between man and machine: Implications of technology for growth, factor shares, and employment},
  author={Acemoglu, Daron and Restrepo, Pascual},
  journal={American economic review},
  volume={108},
  number={6},
  pages={1488--1542},
  year={2018},
  publisher={American Economic Association 2014 Broadway, Suite 305, Nashville, TN 37203}
}

@book{aghion2017artificial,
  title={Artificial intelligence and economic growth},
  author={Aghion, Philippe and Jones, Benjamin F and Jones, Charles I},
  volume={23928},
  year={2017},
  publisher={National Bureau of Economic Research Cambridge, MA}
}

@book{rudin1964principles,
  title={Principles of Mathematical Analysis},
  author={Rudin, Walter},
  year={1964},
  edition={2},
  publisher={McGraw-Hill}
}

@book{tirole1988theory,
  title={The theory of industrial organization},
  author={Tirole, Jean},
  year={1988},
  publisher={MIT press}
}

@techreport{dafoe2018ai,
  title       = {AI Governance: A Research Agenda},
  author      = {Dafoe, Allan},
  institution = {Centre for the Governance of AI, Future of Humanity Institute, University of Oxford},
  year        = {2018},
  month       = aug,
  note        = {Version 1.0, August 27, 2018},
  url         = {https://cdn.governance.ai/GovAI-Research-Agenda.pdf}
}

@misc{openai_charter,
  title        = {OpenAI Charter},
  author       = {{OpenAI}},
  year         = {2018},
  url          = {https://openai.com/charter/},
  note         = {Accessed 2026-04-29}
}

@misc{openai_agi_beyond,
  title        = {Planning for AGI and Beyond},
  author       = {{OpenAI}},
  year         = {2023},
  month        = feb,
  day          = {24},
  url          = {https://openai.com/index/planning-for-agi-and-beyond/},
  note         = {Accessed 2026-04-29}
}

@inproceedings{okeefe2020windfall,
  title        = {The Windfall Clause: Distributing the Benefits of AI for the Common Good},
  author       = {O'Keefe, Cullen and Cihon, Peter and Garfinkel, Ben and Flynn, Carrick and Leung, Jade and Dafoe, Allan},
  booktitle    = {Proceedings of the 2020 AAAI/ACM Conference on AI, Ethics, and Society},
  series       = {AIES '20},
  year         = {2020},
  pages        = {327--331},
  publisher    = {Association for Computing Machinery},
  address      = {New York, NY, USA},
  doi          = {10.1145/3375627.3375842},
  url          = {https://doi.org/10.1145/3375627.3375842}
}

@misc{anderljung2023frontier,
  title        = {Frontier AI Regulation: Managing Emerging Risks to Public Safety},
  author       = {Anderljung, Markus and Barnhart, Joslyn and Korinek, Anton and Leung, Jade and O'Keefe, Cullen and Whittlestone, Jess and Avin, Shahar and Brundage, Miles and Bullock, Justin and Cass-Beggs, Duncan and Chang, Ben and Collins, Tantum and Fist, Tim and Hadfield, Gillian and Hayes, Alan and Ho, Lewis and Hooker, Sara and Horvitz, Eric and Kolt, Noam and Schuett, Jonas and Shavit, Yonadav and Siddarth, Divya and Trager, Robert and Wolf, Kevin},
  year         = {2023},
  eprint       = {2307.03718},
  archivePrefix = {arXiv},
  primaryClass = {cs.CY},
  url          = {https://arxiv.org/abs/2307.03718}
}

@article{ramsey1928mathematical,
  title={A mathematical theory of saving},
  author={Ramsey, Frank Plumpton},
  journal={The economic journal},
  volume={38},
  number={152},
  pages={543--559},
  year={1928},
  publisher={Oxford University Press Oxford, UK}
}

@article{cass1965optimum,
  title={Optimum growth in an aggregative model of capital accumulation},
  author={Cass, David},
  journal={The Review of economic studies},
  volume={32},
  number={3},
  pages={233--240},
  year={1965},
  publisher={Wiley-Blackwell}
}

@incollection{koopmans1965concept,
  author    = {Koopmans, Tjalling C.},
  title     = {On the Concept of Optimal Economic Growth},
  booktitle = {Study Week on the Econometric Approach to Development Planning},
  series    = {Pontificiae Academiae Scientiarum Scripta Varia},
  volume    = {28},
  pages     = {225--287},
  year      = {1965},
  publisher = {North-Holland Publishing Co.},
  address   = {Amsterdam}
}

\ifwithappendix
\clearpage
\appendix
\section{Technical Appendix}
\label{sec:proofs}
Note that in the simulations, we convert the BLS non-farm-business multifactor-productivity index~\citep{BLS_MFPNFBS_2025}, reported as ``index 2017 = 100'', to level form: the 2024 reading of 106.847 becomes $A_{0}=106.847/100\approx 1.068$.
We then project Hicks-neutral productivity forward via  
$A_{t}=A_{0}\,e^{\,g\,(t-2024)}$.
\subsection{Proof of Proposition~\ref{prop:threshold}}
\begin{proof}
\textbf{1. Capital income share.}
Let $r_t$ denote the marginal product of capital, $r_t = \partial Y_t/\partial K_t$, then the capital income share becomes
\begin{equation*}
R(\gamma_t)
     :=\frac{r_tK_t}{Y_t}
     =\gamma_t^{\,1-\rho}\,\bar\alpha^{\,1-\rho}\,
      A_t^{\,\rho}\,
      \left(\frac{K_t}{Y_t}\right)^{\rho},
\end{equation*}
which is strictly increasing in $\gamma_t$ because $1-\rho=1/\sigma>0$.

\textbf{2. Government budget.}
Net public rent equals
$\Theta(1-c)\,R(\gamma_t)Y_t$,
i.e. the government captures a constant fraction of the capital income share defined above.
Balanced transfers therefore require
\begin{equation*}
  \frac{B}{Y_t}
  =\Theta(1-c)\,
   \gamma_t^{\,1-\rho}\,
   \bar\alpha^{\,1-\rho}\,
   A_t^{\,\rho}\,
   \left(\frac{K_t}{Y_t}\right)^{\rho}.
\end{equation*}
Solving for $\gamma_t^{\,1-\rho}$ and raising to
$\sigma=1/(1-\rho)$ gives the desired $\gamma_t^{\star}$.
Strict monotonicity of $R$ with respect to $\gamma_t$ (as it is of the form $R(\gamma_t) \equiv C_t\gamma_t^{1-\rho}$ for $C_t > 0$) guarantees uniqueness~\citep{rudin1964principles}.

\textbf{3. BGP existence and convergence.}
With constant saving $K_{t+1}=sY_t+(1-\delta)K_t$, the ratio $q_t := K_t/Y_t$ satisfies $q_{t+1}=F(q_t)$ for a continuous, increasing map $F$ with a single fixed point $q^\ast=\bar\kappa = s/(e^g - 1 +\delta)$.
Since the update map $F(q)=\dfrac{s+(1-\delta)q}{e^{g}}$ is strictly increasing and a contraction because $F'(q)=(1-\delta)/e^{g}<1$, the Banach fixed-point theorem~\citep{rudin1964principles} yields global (linear) convergence $q_t\to q^\ast$ for any initial $q_0$.
Hence, if capability never falls below the threshold, $\gamma_t \ge \gamma_t^{\star}$ for all $t$, then net public rent $\Theta(1-c)R(\gamma_t)Y_t$ weakly exceeds $B$ each period, so the transfer stays solvent along the whole (convergent) path.
\end{proof}

\subsection{Proof of Corollary~\ref{cor:comparative}}
\begin{proof}
Let $\ell_t:=\ln\gamma^{\star}_t=\sigma\ln Z_t$.  
Holding $B/Y_t,\bar\alpha,A_t$ fixed,
\begin{equation*}
\begin{split}
&\partial_B \ell_t = \frac{\sigma}{B},\qquad \partial_{\delta}\ell_t = \frac{\sigma-1}{e^g - 1 + \delta},\qquad \partial_{\bar\alpha}\ell_t = -\frac{1}{\bar\alpha},\\
&\partial_\Theta\ell_t=-\frac{\sigma}{\Theta},\qquad
\partial_c\ell_t=\frac{\sigma}{1-c},\qquad
\partial_s\ell_t=-\frac{\sigma\rho}{s},\\
&\partial_{\bar\kappa}\ell_t = -\frac{\sigma\rho}{\bar\kappa},\qquad \partial_\sigma\ell_t=\ln Z_t+\frac{1}{\sigma}\ln\left(\dfrac{\bar\alpha}{A_t\bar\kappa}\right).
\end{split}
\end{equation*}
Because $d\gamma^{\star}_t/d\ell_t = \gamma^{\star}_t$, multiplying each line by $\gamma^{\star}_t$ gives the stated elasticities.
\end{proof}

\subsection{Proof of Proposition~\ref{prop:oligopoly}}
\begin{proof}
\emph{Profit share.} Cournot first-order conditions~\citep{tirole1988theory} imply $(P-\mathrm{MC})/P=\theta/\eps$, so pure profit (denoted by $\Pi_t$) accounts for the fraction $\Pi_t/Y_t=\theta/\eps$ of output.

\emph{Budget constraint.} Government income equals
\begin{equation*}
\Theta(1-c)\Bigl[R(\gamma_t)+\tfrac{\theta}{\eps}\Bigr]Y_t,
\qquad
R(\gamma_t):=\gamma_t^{1-\rho}\bar\alpha^{1-\rho}A_t^{\rho}{\bar\kappa}^{\rho}.
\end{equation*}
Setting this $\ge B$ and solving for $\gamma_t$ yields the boxed
$\gamma^{\star}_{\mathrm{oligo},t}$.

\emph{Competitive benchmark.} Under perfect competition, $\theta=0$, recovers $\gamma^{\star}_{\mathrm{comp},t} =
   \bigl(\tfrac{B/Y_t}{\Theta(1-c)\bar\alpha^{1-\rho}A_t^{\rho}\bar\kappa^{\rho}}\bigr)^{\sigma}$.
Since subtracting $\theta/\eps>0$ in the numerator lowers the fraction inside the parentheses and $\sigma>0$, the oligopoly threshold is strictly smaller whenever $\theta>0$ (imperfect competition).
\end{proof}

\subsection{Proof of Corollary~\ref{cor:theta}}
\begin{proof}
Factor the common constant $C_t$ out of both thresholds:
\begin{equation*}
\gamma^{\star}_{1,t}=C_t\,\Theta_1^{-\sigma},
\qquad
\gamma^{\star}_{2,t}=C_t\,\Theta_2^{-\sigma}.
\end{equation*}
Because $\sigma>0$ and $\Theta_2>\Theta_1$, we have
$\Theta_2^{-\sigma}<\Theta_1^{-\sigma}$, implying
\(\gamma^{\star}_{1,t}>\gamma^{\star}_{2,t}\).
Subtracting confirms the boxed expression and its strict
positivity.
\end{proof}

\subsection{Endogenous Saving Does Not Change the AI Capability Threshold Logic}
\label{ss:endo}
In our baseline Solow-Zeira framework we assume an exogenous saving rate $s$.
This is a deliberate simplification: our closed-form AI capability threshold $\gamma_t^\star$ rises with the saving rate (cf. Corollary~\ref{cor:comparative}), so any reduction in the effective saving rate will lower the capability required to finance a universal basic income.
However, an endogenous saving rate can be studied in a Ramsey-Cass-Koopmans framework~\citep{ramsey1928mathematical,cass1965optimum,koopmans1965concept}.
There a representative agent chooses consumption $C_t$ to maximize $\sum_{t=0}^{\infty}\beta^t U(C_t)$ subject to the per‑worker resource constraint
\begin{equation*}
  C_t + K_{t+1} = f(K_t) + (1-\delta)K_t,
\end{equation*}
where $f(K_t)$ is the per‑worker CES production function derived from our Solow-Zeira aggregator in equation~\eqref{eq:ai-ces} with labor $L$ normalized to one:
\begin{equation*}
  f(K_t) \;\equiv\; A_t\left(\gamma_t^{1-\rho}\,\bar{\alpha}^{\,1-\rho}\,K_t^{\rho} \;+\; (1-\bar{\alpha})^{1-\rho}\right)^{1/\rho}.
\end{equation*}
The first‑order conditions deliver the Euler equation 
\begin{equation*}
  U'(C_t) = \beta\,[1 - \delta + f'(K_{t+1})]\,U'(C_{t+1}),
\end{equation*}
which, in steady state $C_{t+1} = C_t$ and $K_{t+1} = K_t = K^\star$, the Euler equation simplifies to
\begin{equation*}
  1 = \beta\,[1 - \delta + f'(K^\star)] \quad\Longleftrightarrow\quad f'(K^\star) = \frac{1}{\beta} - (1-\delta).
\end{equation*}
By concavity of $f$, $f'(K^\star) \le f(K^\star)/K^\star = 1/\kappa_{\mathrm{end}}$, where $\kappa_{\mathrm{end}} = K^\star/Y^\star = K^\star/f(K^\star)$ is the Ramsey steady‑state capital-output ratio.
Therefore, if
\begin{equation*}
\bar\kappa = \frac{s}{e^g - 1 + \delta} > \frac{1}{\frac{1}{\beta} - (1-\delta)},
\end{equation*}
then $\bar{\kappa} > \kappa_{\mathrm{end}}$.
In our calibration in \S\ref{sec:calibration}, this strict ordering $\bar\kappa>\kappa_{\mathrm{end}}$ holds for sufficiently impatient discount factors ($\beta\lesssim0.8$), but the comparative static of Corollary~\ref{cor:comparative} remains: reducing savings $s$ \emph{always} reduces $\gamma_t^\star$---whether induced by transfers, heterogeneity, or impatience.  
Moreover, exogenous saving-rate models continue to serve as tractable benchmarks; for example, \citet{xepapadeas2016spatial} analyze a spatial growth model where saving rates are fixed exogenously to obtain tractable analytical results.
Our framework follows this tradition: fixing the net saving rate simplifies the algebra, delivers a closed-form threshold, and typically errs on the side of overestimating the capability needed to finance broad-based benefits.
\fi
\end{document}